\documentclass[aps]{revtex4}
\usepackage{graphicx}

\begin{document}

\hfill APPLIED PHYSICS REPORT 2004--17

\vspace{2cm}

\title{van der Waals interactions of the benzene dimer: 
towards treatment of polycyclic aromatic hydrocarbon dimers}

\author{Svetla D.~Chakarova}\thanks{Corresponding author: Tel. +46 31 772 3632;
Fax +46 31 772 8426; E-mail \texttt{svetla@fy.chalmers.se}}
\author{Elsebeth Schr\"{o}der}

\affiliation{Department of Applied Physics, Chalmers University of Technology and
G\"{o}teborg University, SE--412 96 G\"{o}teborg, Sweden}

\date{June 21, 2004}

\begin{abstract}

Although density functional theory (DFT) in principle includes
even long-range interactions, standard implementations employ
local or semi-local approximations of the interaction energy and
fail at describing the van der Waals interactions. We show how to
modify a recent density functional that includes van der Waals
interactions in planar systems [Phys.~Rev.~Lett.~\textbf{91}, 126402
(2003)] to also
give an approximate interaction description of planar molecules. 
As a test case we use this modified
functional to calculate the binding distance and energy for
benzene dimers, with the perspective of treating also larger,
flat molecules, such as the polycyclic aromatic hydrocarbons (PAH).\\

\end{abstract}

\maketitle


\section{Introduction}

The benzene dimer is the prototype for aromatic interactions and has
been studied extensively both by theoretical 
\cite{theo1,theo2,theo3,theo4,theo5,theo6} 
and experimental \cite{exp1,exp2,exp3}
means. Describing benzene interactions can be regarded as the first step toward
describing interactions of polycyclic aromatic hydrocarbons (PAHs). 
PAHs are planar molecules consisting of
several aromatic rings, where the peripheral carbon atoms are
covalently bonded to hydrogen atoms. Both benzene and PAHs 
structurally resemble graphite and exhibit very similar 
intra- and intermolecular bond lengths, particularly for 
(stacks of) large PAHs.
Like the interactions between the sheets in graphite, the
interactions between parallel benzene molecules or layers in PAH stacks are 
dominated by the weak, nonlocal van der Waals (vdW) interaction. 
However, the description of the vdW interaction is absent in the 
traditional implementations 
of density functional theory (DFT), implementations which have otherwise 
been  very successful in describing dense, hard materials on the 
atomic scale. 

The past few years a number of publications 
\cite{kohnvdW,Unified,dobson,tractable,bostrom,layerprl,ijqc,misquitta,ggpaper}
have addressed the problem of consistently extending the common 
DFT implementations to also include the vdW interaction.
For the mutual interactions of planar sheets or surfaces a functional 
was obtained in Refs.\ \cite{tractable,layerprl}. 
For example, a relatively good estimate of the binding distance between 
two graphite planes (graphene) was obtained with this planar 
functional \cite{layerprl}, while the usual 
generalized gradient approximations (GGA) fail to bind the graphene
layers or do so at an unphysical binding distance and energy.
Since benzene and PAHs are similar to graphite, the correction to the regular
calculations will be significant also in dimers of these molecules.

In this work, we study dimers of benzene (C$_{6}$H$_6$)
by modifying and approximating the planar nonlocal density
functional \cite{tractable,layerprl}
in order to treat molecules of finite size.
We here concentrate on dimers where the molecules are placed directly on
top of each other (``AA stacking'').
The method we use is developed 
with the aim of being able to study also PAH molecules,
and it is described below. The results for some of the small PAH
molecules   naphthalene, anthracene, and pyrene (C$_{10}$H$_8$,
C$_{14}$H$_{10}$, and C$_{16}$H$_{10}$)
will be reported in a forthcoming publication \cite{svetlapah}.

\section{Method}

DFT in principle does include even long-range interactions, such as
the vdW interactions, but those are a part of the exchange-correlation
functional $E_{xc}$ which is approximated in the implementations of DFT. 
The most commonly used approximations to $E_{xc}$ are
the local density approximation (LDA) and the generalized gradient
approximation (GGA), both of which are local or semi-local and do
not include the vdW interaction. However, GGA (and to some extend LDA)
has proven very successful in describing the interactions within
dense matter, e.g., within a molecule, and a consistent
correction scheme that includes the long-range intermolecular
vdW interactions must not
change the short-range intramolecular interaction. 

\subsection{The graphite interactions}

In the vdW correction scheme for planar sheets and surfaces, 
defined and described in Refs.\ \cite{tractable,layerprl,ijqc},  
the exchange energy from GGA is retained, and the correlation
energy is determined as a sum of the local correlation (obtained from LDA)
and the nonlocal correlation $E^{nl}_c$, which will be described below.
For the exchange part the GGA revPBE flavor of Zhang and Yang \cite{zhang} 
is chosen because it is fitted to exact exchange for atoms.
Thus the (vdW-corrected) total energy is written as
\begin{equation}
  E_{\mathrm{vdW-DF}} =  
  E_{\mathrm{GGA}}- E_{\mathrm{GGA},c} + E_{\mathrm{LDA},c} + E^{nl}_c,
  \label{eq:vdW-DF}
\end{equation}
where all energy terms are functionals of the electron charge density
$n(\mathbf {r})$, and $E_{\mathrm{GGA},c}$ and $E_{\mathrm{LDA},c}$ 
are the correlation from GGA and LDA, respectively. These energy
terms are directly available from our DFT calculations.
Our DFT calculations are performed by the plane-wave pseudo-potential
based program Dacapo \cite{dacapokod}.

The nonlocal correlation, per area $A$, for a planar, translationally invariant
system such as graphite is given by \cite{tractable}
\begin{equation}
E^{nl}_c/A= -\lim_{L\rightarrow\infty} \int_0^\infty \frac{du}{2\pi}
\int\frac{d^2 k}{(2\pi)^2} \ln \frac{\phi'(0)}{\phi'_0(0)}
\label{eq:ecnl}
\end{equation}
where $\phi(z)$ fulfills the differential equation 
$(\epsilon_k\phi')'=k^2 \epsilon_k^2 \phi  $ with boundary conditions
$\phi(0)=0$ and $\phi(L)=0$, and $\phi_0(z)$ is the vacuum solution.
The system is enclosed in a long box of length $L$ along the direction 
perpendicular to the graphite plane(s).
$\epsilon_k$ is the dielectric function of
graphite, Fourier-transformed in the plane of the graphite sheet
\begin{equation}
  \epsilon_k(z, iu)=1+ \frac {\omega^2_p(z)}{u^2+ 
  \nu_F^2 (z)(k^2+q_{\perp}^2)^2/3 + (k^2+q_{\perp}^2)^2/4}\,.
  \label{eq:dielmodel}
\end{equation}
The plasmon frequency
$\omega _p$ is defined by $\omega _p^2 (z) = 4 \pi n(z) $, 
the Fermi velocity by $ \nu_F (z) = (3 \pi ^2 n(z))^{\frac {1}{3}}$, 
$n(z)$ is the planarly averaged electron density varying in the
$z$-direction perpendicular to the plane,
and $iu$ is the imaginary frequency. We use atomic units unless otherwise
noted.

The parameter $q_{\perp}$ is introduced to compensate for the locality
of $\epsilon_k$ in the $z$-direction. The value of $q_{\perp}$ is 
materials dependent, and it is found
from a separate set of GGA DFT calculations with an applied, static electric
field across the graphite layer. 

In Figure \ref{fig:tot2}(a) we show the  interaction
curve of graphene-graphene \textit{in the
AA stacking\/} according to (\ref{eq:vdW-DF}). These results of  
graphene in the AA stacking allow us to directly 
compare to our results of benzene, and later PAHs, also in AA stacking.
The binding separation 4.0 {\AA} of 
graphene (AA) is larger than for graphene in the physically 
correct AB stacking (3.76 {\AA}), calculated within the same vdW correction
scheme \cite{layerprl}.
We used the value $q_{\perp}= 0.756$ in atomic units for a graphene
sheet \cite{layerprl}.

\subsection{Extending the functional to treat planar molecules}

It is a natural next step to modify $E^{nl}_c$ for
planar, translational invariant sheets to treat planar, large,
but not infinitely extended molecules. The PAH molecules 
are pieces of graphene,
passivated by hydrogen atoms at the broken carbon-carbon bonds. 
We would therefore like to introduce appropriate approximations 
in order to apply the functional to planar, large, finite molecules.

The (semi-)local part of the total energy in (\ref{eq:vdW-DF}),
$E_{\mathrm{GGA}}- E_{\mathrm{GGA},c} + E_{\mathrm{LDA},c}$, 
can  be determined directly from a usual 
DFT implementation. Such calculations provide us, besides the energy 
terms, with a three-dimensional electron density $n(\mathbf{r})$. 
In the formalism for $E^{nl}_c$ in (\ref{eq:ecnl}) the density profile 
$n(z)$ is assumed translationally invariant along the molecule.
For the finite molecules  we take 
$n(z)= A_{\mathrm{mol}}^{-1}\int dxdy\, n(x,y,z)$ 
where $A_{\mathrm{mol}}$ is the size of the molecular area that affects 
the vdW interactions; this area must be estimated.
%$x$ and $y$ are coordinates parallel to the molecule.

For carbon-dominated molecules such as benzene and PAH the dominating 
contribution to the vdW correction comes from the carbon atoms.
We thus assume that the dielectric function for the molecules 
is well described by the functional form (\ref{eq:dielmodel})
with the graphene value $q_{\perp}=0.756$ a.u. By comparing
the dacapo-calculated static ($u=0$) electric response of the 
isolated molecule with that found from the model dielectric function 
(\ref{eq:dielmodel}), with $A_{\mathrm{mol}}$ as a parameter, we can determine
$A_{\mathrm{mol}}$. For benzene we find that $A_{\mathrm{mol}}=51.7$ {\AA}$^2$
reproduces the static response. 

Given the model dielectric function (\ref{eq:dielmodel}) with the
parameter $A_{\mathrm{mol}}$ determined as described above, the 
calculation of the $E^{nl}_c$ contribution follows from (\ref{eq:ecnl}).
The physical analogy of using the molecularly averaged $n(z)$ in
(\ref{eq:ecnl}) is that of molecules moved together such that
they each take up the lateral area $A_{\mathrm{mol}}$.
$E^{nl}_c$ for molecules calculated this way therefore somewhat overcounts 
the vdW interaction by including not only the physically
correct interaction of one dimer molecule with its partner, but
also the interaction with the partner's neighbors. 
Because the vdW attraction falls off with distance this effect
decreases as the molecule size increases.
The other energy terms of (\ref{eq:vdW-DF}) are not affected.

We stress that $A_{\mathrm{mol}}$ is larger than the 
area of the same amount of carbon atoms in a
graphene plane. In our procedure of finding $A_{\mathrm{mol}}$
we compare the (static) response of a molecule isolated
in all three directions. We make sure that our Dacapo-calculated 
response really does
come from an isolated molecule by checking the convergence
of the response with increasing lateral unit cell size.
The response of the low-density charge distribution region within the 
molecular plane is thus included in our description, and
therefore the vdW interaction of these sheets of close molecules
is different than the vdW interaction of two sheets of graphite,
even though the molecules are mostly composed of
carbon atoms. 

With the assumptions introduced above, necessary for 
applying the vdW planar functional to finite molecules,
the quality of the vdW interaction calculations increases with 
increasing molecular size. We emphasize that in
this sense, the benzene molecule is an extreme molecule, and we do not
expect full quantitative agreement of our results with
full-powered quantum-chemistry calculations \cite{theo1,theo6} or with results
in our new (more costly) general-geometry formulation of the
vdW-DF \cite{ggpaper}.

\section{Results}

We here present the results of applying the modified 
vdW functional to the benzene dimer in the AA stacking.
The first step is to determine the structure of the isolated
molecule, the static response to an applied electric field,
the (semi-)local energy contributions to the
dimer binding, and the electron density of the dimer.
This step is carried out in a standard DFT program.
The second step is to calculate the nonlocal correlation
energy contribution, $E^{nl}_c$, with the approximations
discussed above.

The standard DFT calculations were performed with the
plane-wave pseudopotential code Dacapo \cite{dacapokod}
with periodic boundary conditions. We performed the
self-consistent calculations (before adding $E^{nl}_c$) 
using the revPBE GGA functional.
The choice of  unit cell size involves a trade-off between manageable
calculations on the one hand, and large
lateral and vertical separations between periodically repeated images of
the molecules (or dimers) 
on the other hand. When deciding on the unit cell size we also benefit
from the short-range nature of the traditional DFT-implementations.
In our calculations we used a hexagonal unit cell of size (17.112 \AA,
17.112 {\AA}, 26 {\AA}). 

In the plane-wave code the number of plane 
waves is determined by a cutoff in the plane-wave energy. 
A preliminary analysis showed a cutoff at 450 eV to be
sufficient. However, as we discuss below, there is a small 
energy difference between dimers within the same unit cell
at very large separation
and two isolated molecules in two separate unit cells.
This offset depends on the 
plane-wave cutoff in a non-trivial way. We emphasize that the
offset in the total energy difference is 
%present already \textit{without\/} 
%adding $E^{nl}_c$, and is 
unrelated to $E^{nl}_c$.

\subsection{Benzene-benzene interaction energy}

\begin{figure}
  \begin{center}
    \includegraphics[width=8cm]{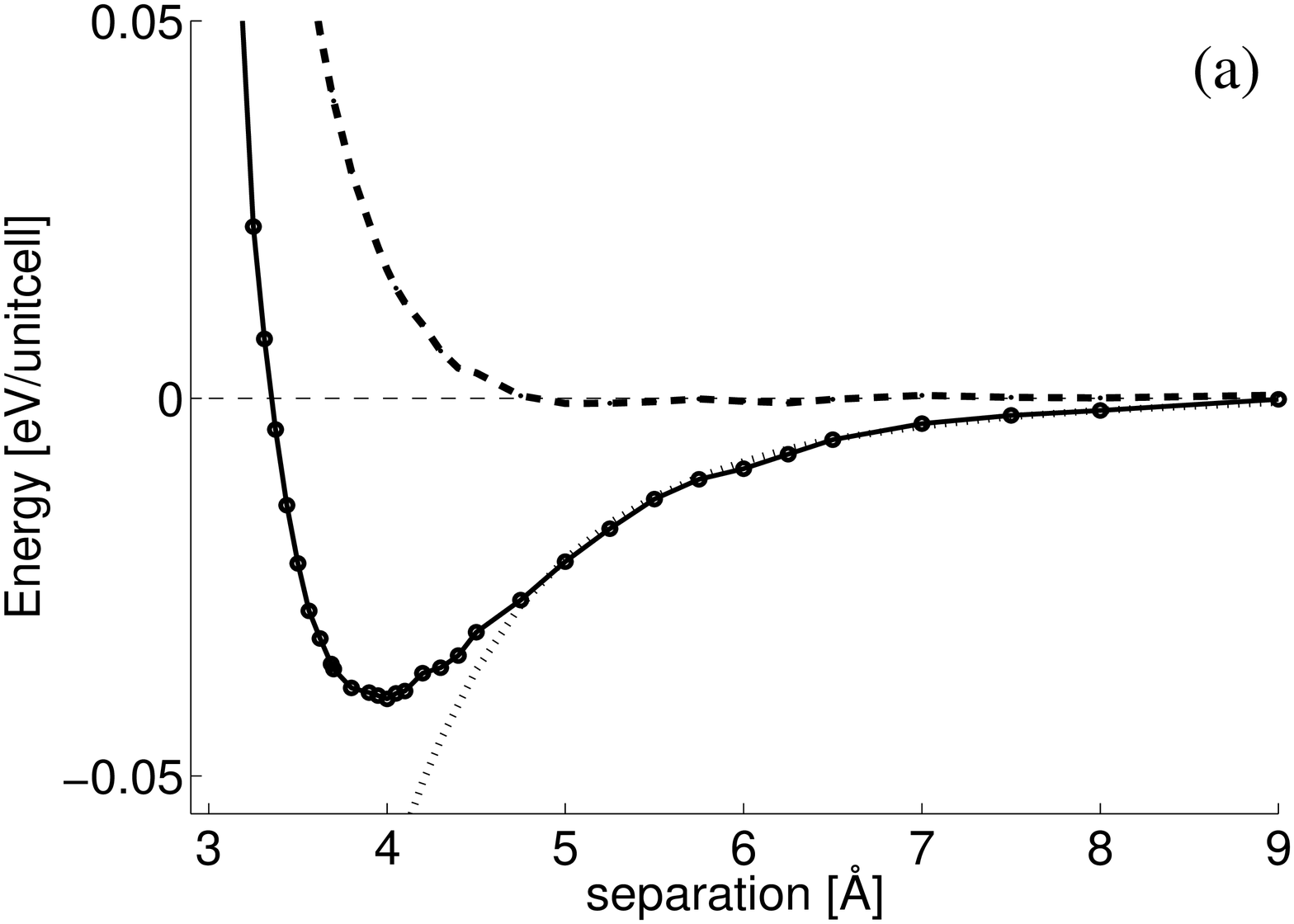}\\[1em]
    \mbox{\hspace{2mm}}\includegraphics[width=8cm]{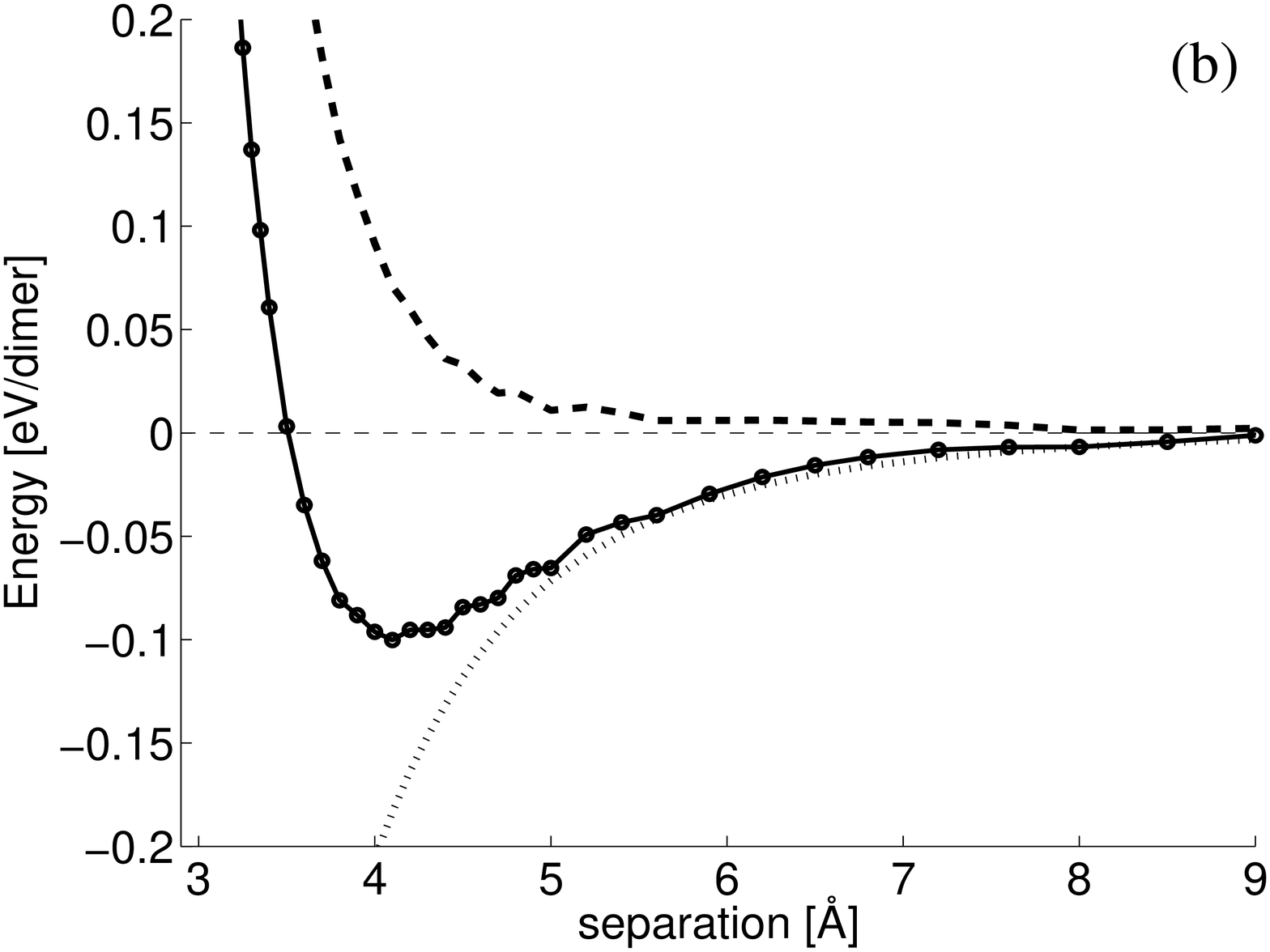}
  \end{center}  
  \caption{Energy curves for graphene and benzene dimers in the AA stacking as
a function of separation distance.
   Panel (a) shows the graphene-graphene interaction, panel (b) 
shows the benzene-benzene interaction.  
   The dashed lines are the self-consistently determined revPBE total energies, 
   the dotted lines show the $E_c^{nl}$ correction, 
   and the solid lines show the $E_{\mathrm{vdW-DF}}$ total
   energy curves obtained according to Eq.\ (\protect\ref{eq:vdW-DF}).
   The circles on the $E_{\mathrm{vdW-DF}}$ curves indicate the calculations
performed.}{ \label{fig:tot2}}
\end{figure}

The benzene dimer interaction is calculated by fixing the relative atomic %relaxed
positions within of each of the two molecules of the dimer and
varying the distance between them, while keeping the size of the unit cell
constant. 
Figure \ref{fig:tot2}(b) shows the total-energy curves for benzene: the
revPBE curve (which shows no binding at all), the nonlocal correlation $E^{nl}_c$,
and the full $E_{\mathrm{vdW-DF}}$ that includes nonlocal correlation.
The reference energy is the benzene dimer at large (11 {\AA}) separation. 

We find in vdW-DF the binding energy $100$ meV at the distance 4.1 {\AA}. 
These results are in agreement with our results for the AA 
stacked graphene dimer as we expect the binding distance there to be similar 
to that of the benzene dimer and PAH dimers in the AA stacking. 
Both the benzene separation distance and interaction energy are in
reasonable, although not perfect
agreement with experiments and other calculations. For instance, 
Refs.\ \cite{theo2,theo4,theo5} report calculating binding 
separations of around 3.8 {\AA} and binding energies in the range of 64--92 meV.
As mentioned earlier, we expect the quality of our
results to improve for larger molecules.

As previously mentioned, no binding arises in the revPBE flavor of GGA.
We note that GGA Perdew-Wang 91 (PW91) \cite{PW91} calculations (not shown) exhibit 
an unphysically small binding energy of the benzene dimer ($12$ meV) at 
separation 4.7 {\AA}. However, this PW91 binding arises from the unphysical
interactions mediated by the exchange contribution  
in the manner previously documented for graphite \cite{layerprl}.

\subsection{Discussion of the energy reference level}

During the course of this benzene and PAH study we encountered
a technical problem in determining an appropriate reference
level for our (traditional) total energy calculations based on GGA. 

All dimer energies presented in this paper, whether revPBE or vdW-DF 
energies, are given with reference to the energy of two 
molecules far apart. Thus the energies that we report, and which we
for now generally denote $E_{\mathrm{diff}}$, are total energy 
\textit{differences} given by 
\begin{equation}
  E_{\mathrm{diff}} (d)= E_{\mathrm{tot}}(d)- E_{\mathrm{ref}}
  \label{eq:dimere_d}
\end{equation}
where $E_{\mathrm{tot}}(d)$ is the total energy of the dimer determined
at separation $d$, and $E_{\mathrm{ref}}$ is the energy of
two molecules far apart.
This applies both to the benzene molecules and the
graphite sheets. The reference level $E_{\mathrm{ref}}$
can be calculated either by removing one molecule to find half the reference energy,
or by keeping both molecules
within the unit cell and moving them ``far apart'' within the box.
In both cases, we make sure that
the periodic cell is sufficiently large that the molecule does not
interact (electrostatically)
with its periodic images. In our underlying GGA-DFT
calculations it turned out that 
these two seemingly equivalent ways of determining $E_{\mathrm{ref}}$
lead to two slightly different results.  Numerically, the difference
is more significant for the benzene molecules than for
graphene. 

In Fig.~\ref{fig:tot2}(b)
 we report curves of interaction energies of
dimers with increasing separation. The natural choice of reference
energy $E_{\mathrm{ref}}$ is therefore the dimer with the two molecules 
``far apart'', by which we mean 11--12 {\AA} separation in  boxes 
of length of 24--26 {\AA}. However, the problem merits an analysis.

\begin{figure}
  \begin{center}
    \includegraphics[height=8cm]{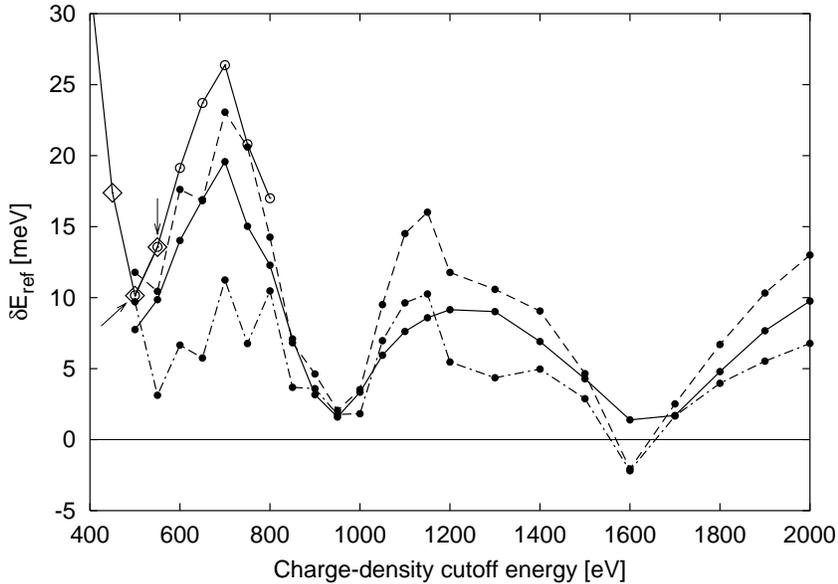}
  \end{center}
  \caption{Difference in reference energies $\delta E_{\mathrm{ref}}$ as
a function of charge-density cutoff energy. Diamonds are for
calculations with wave-function cutoff energy equal to the charge-density
cutoff energy, empty circles have wave-function cutoff 450 eV in the usual unit cell
size and black circles have wave-function cutoff 450 eV in a 
 laterally smaller unit cell, as described 
in the text. Solid, dashed, and dash-dotted lines show $\delta E_{\mathrm{ref}}$ 
of the revPBE
total energy, of revPBE exchange, and of PW91 exchange.}
 { \label{fig:offset}}
\end{figure}

In Figure \ref{fig:offset} we show the difference in reference
energies $\delta E_{\mathrm{ref}}$ --- defined as the energy of a pair of molecules 
``far apart'' subtracted by the energy of two separate molecules
--- as a function of the choice of DFT convergence parameters. 
In Dacapo, besides the plane-wave energy cutoff for the wave functions, it is possible 
to set a different (larger) cutoff for the charge density, to enhance
the description of the charge density at the price of only a modest increase in 
the calculational size.
Most of the calculations in Figure \ref{fig:offset} use a larger 
charge-density cutoff than the wave-function cutoff. 
In order to analyze $\delta E_{\mathrm{ref}}$ also for large charge-density
cutoff energies we carried out a number of calculations in a unit cell
with reduced lateral size (10 {\AA}).

A few immediate conclusions can be drawn on the basis of Figure \ref{fig:offset}.
By comparing the single-cutoff and the double-cutoff results at charge-density
cutoff 500 eV and 550 eV we find that the quality of the wave-function 
description does not affect $\delta E_{\mathrm{ref}}$. We see this
in Figure \ref{fig:offset} by the collapsed data points  
(indicated by arrows) for the revPBE total energy, but it is
true also for the exchange and the correlation parts of the energy, as well
as for PW91 (not shown). This indicates that the (noninteracting) 
kinetic part of the total energy, which is calculated from the wave functions,
does not contribute to $\delta E_{\mathrm{ref}}$.

In contrast, $\delta E_{\mathrm{ref}}$ depends
non-trivially on the charge-density cutoff. It is not clear that a
very high cutoff value will yield good results, rather, $\delta E_{\mathrm{ref}}$
seems to oscillate with charge-density cutoff.
By comparing the results for the usual box size 
with the small-box results we further notice that $\delta E_{\mathrm{ref}}$
decreases with the amount of low-density (vacuum) region in the box. 
$\delta E_{\mathrm{ref}}$ thus seems to be mainly affected by contributions 
from the low-density regions. A major part of $\delta E_{\mathrm{ref}}$
comes from the exchange part of the total energy. Not shown in the figure
are $\delta E_{\mathrm{ref}}$ for
the PW91 total energy and for correlation (revPBE and PW91). These
energies all give a finite, but smaller value of $\delta E_{\mathrm{ref}}$.
Unfortunately the problems are worse exactly for the
revPBE-GGA that enters our subsequent vdW-DF calculations 
in (\ref{eq:vdW-DF}).

Besides the ever-present risk of undetected programming errors,
we speculate on three possible origins for the finite $\delta E_{\mathrm{ref}}$
that affects our underlying GGA-DFT calculations:
(i) numerical instabilities and formal inaccuracies arising from
the implementation of the exchange-correlation approximation; 
(ii) the mixing of different 
exchange-correlation approximations for creation of pseudopotentials and 
use in calculations; and 
(iii) the use and/or implementation of ultrasoft pseudopotentials.
The possibility of formal problems in the exchange-correlation term
borrows some
support from the work by Lacks and Gordon in 1993 \cite{lacks}:
PW91 exchange, and a number of other exchange approximations,
were shown to give values of $|\delta E_{\mathrm{ref}}|$
for noble gas dimers far exceeding those from
exact exchange. The deviation arises from contributions
in the low-density, large-gradient regions of space, which is
precisely the regions that dominate our system. 
If this will turn out to be the origin \cite{footnote}.
%\footnote[1]{\baselineskip 24pt The (semi-)local nature of
%GGA ensures that GGA implicitly
%accounts also for part of the so-called 
%``static'', or non-dynamical, correlation.
%This, however, is a constant value.

%Since neither of the two correlation terms $E_{\mathrm{LDA},c}$
%or $E_c^{nl}$ include static 
%correlation we must for vdW-DF use an exchange term that mimics static 
%correlation in addition to exact exchange.
%}
 
of at least a part of the offset $\delta E_{\mathrm{ref}}$,
it is not too surprising that this does not seem to 
be perceived as a general problem, since 
GGA and most present-day DFT codes have been 
implemented and tested mostly on dense systems 
with less vacuum.

We might worry that using a mix of exchange-correlation
approximations in the pseudo{\-}potential and in the energy calculations
could lead to unwanted effects. 
Our calculations within traditional DFT were carried out
with Vanderbilt ultrasoft pseudopotentials \cite{vanderbilt},  
originally created within the PW91 flavor of the GGA functional, and
the energy and the charge density were self-consistently determined
within revPBE. However, we carried out
a set of calculations for the graphene-graphene system, where 
the offset is much less in size, that showed
this mixing to not be the cause of the offset. 
The calculations of $\delta E_{\mathrm{ref}}$
were carried out with
all combinations of pseudopotentials (revPBE, PW91, PBE \cite{pbe}),
self-consistently determined charge density (revPBE, PW91, PBE),
and energy extracted (revPBE, PW91, PBE, RPBE \cite{rpbe}, LDA, as well as all
exchange and correlation contributions). In all cases, the major part
of the offset arises from the exchange part of the energy, whereas
the offset on the GGA and LDA correlation is generally 
almost an order of magnitude smaller.

We cannot, within this study, test
whether the use of ultrasoft pseudopotentials contributes to 
$\delta E_{\mathrm{ref}}$.
This issue requires controlled comparison of results from different
codes and choice of pseudopotentials and will be addressed in
a forthcoming investigation.

In addition to the tests mentioned above, 
we also tested that the exact position of the monomer
with respect to the Fast-Fourier-Transform grid does not
matter on the scale of energy differences considered here.
Further,  we speculated that $\delta E_{\mathrm{ref}}$ 
might be due to numerical errors arising when very small 
charge-density values are left out in the calculation of the energies,
but by changing the limit of the smallest charge-density values included
we found this to be irrelevant to $\delta E_{\mathrm{ref}}$.

\section{Conclusions}

We have presented a modification of an existing planar-system 
van-der-Waals DFT functional to approximately treat also 
large but finite planar molecules. We use the modified functional
for the benzene dimer as an example, with the prospect of
applying it to polycyclic aromatic hydrocarbon molecules. 
We have characterized  a problem in traditional implementations of DFT 
for molecular systems with large
volumes of low charge density. This problem motivates further investigations.
The modified vdW functional presented is a step towards
being able to treat sparse matter systems consistently within DFT.

\section{Acknowledgments}

We wish to thank Aaron Puzder for helpful discussions. This project
was partly supported by the Swedish foundation for strategic research (SSF)
through consortium ATOMICS, by
the EU Human potential research training network
ATOMCAD under contract number HPRN-CT-1999-00048,
and by the Swedish Research Council (VR).
Allocation of computer resourses through the Swedish National
Allocation Committee (SNAC) is gratefully acknowledged.

%\pagebreak
%%%%%%%%% References
%\vspace{-\parskip}
%\thebibliography{99}

%\bibitem{benzenetheory}benzenetheory


\begin{thebibliography}{99}
\bibitem{footnote}
{The (semi-)local nature of GGA ensures that GGA implicitly
accounts also for part of the so-called 
``static'', or non-dynamical, correlation.
This, however, is a constant value.}
\end{thebibliography}
\end{document}